\pdfoutput=1
\documentclass[num-refs]{nbdt-article}

\newcommand{\beginsupplement}{%
        \setcounter{table}{0}
        \renewcommand{\thetable}{S\arabic{table}}%
        \setcounter{figure}{0}
        \renewcommand{\thefigure}{S\arabic{figure}}}

\papertype{Original Article}
\paperfield{Analysis}

\title{A zero-inflated gamma model for deconvolved calcium imaging traces}

\author[1,2\authfn{1}\authfn{2}]{Xue-Xin Wei}
\author[1\authfn{1}]{Ding Zhou}
\author[2]{Andres Grosmark}
\author[3]{Zaki Ajabi}
\author[2]{Fraser Sparks}
\author[1,2]{Pengcheng Zhou}
\author[3]{Mark Brandon}
\author[2]{Attila Losonczy}
\author[1,2]{Liam Paninski}

\contrib[\authfn{1}]{Equally contributing authors.}
\contrib[\authfn{2}]{Present address: Department of Neuroscience, UT Austin.}

\affil[1]{Departments of Statistics, Grossman Center for the Statistics of Mind, Center for Theoretical Neuroscience, Columbia University}
\affil[2]{Department of Neuroscience, Zuckerman Mind Brain Behavior Institute, Columbia University}
\affil[3]{Integrated Program in Neuroscience, Douglas Hospital Research Centre, McGill University}

\fundinginfo{{Supported by NIBIB R01 EB22913, NSF Neuronex DBI-1707398, NIH 1U19NS104649-01 (LP); R01 NS094668, U19 NS104590, R01 MH100631 (AL); Charles H. Revson Foundation Biomedical Fellowship (AG); National Sciences and Engineering Research Council Discovery grant \#74105, Canada Research Chairs Program (MB); the Gatsby Charitable Foundation to the Center for Theoretical Neuroscience at Columbia University.}}

\corraddress{Xue-Xin Wei, Liam Paninski\\ weixx@utexas.edu, liam@stat.columbia.edu}

\runningauthor{Wei, Zhou et al.}

\begin{document}

\maketitle

\begin{abstract}
Calcium imaging is a critical tool for measuring the activity of large neural populations. Much effort has been devoted to developing ``pre-processing'' tools for calcium video data, addressing the important issues of \emph{e.g.}, motion correction, denoising, compression, demixing, and deconvolution.  However, statistical modeling of deconvolved calcium signals (i.e., the estimated activity extracted by a pre-processing pipeline) is just as critical for interpreting calcium measurements, and for incorporating these observations into downstream probabilistic encoding and decoding models. Surprisingly, these issues have to date received significantly less attention. In this work we examine the statistical properties of the deconvolved activity estimates, and compare probabilistic models for these random signals. In particular, we propose a zero-inflated gamma (ZIG) model, which characterizes the calcium responses as a mixture of a gamma distribution and a point mass that serves to model zero responses.  We apply the resulting models to neural encoding and decoding problems.  We find that the ZIG model outperforms simpler models (\emph{e.g.}, Poisson or Bernoulli models) in the context of both simulated and real neural data, and can therefore play a useful role in bridging calcium imaging analysis methods with tools for analyzing activity in large neural populations.
\end{abstract}

\section{Introduction}
Calcium imaging is one of the primary methods for measuring the activities of large neural populations at single-cellular resolution~\cite{Yuste1995,Svoboda1997,Helmchen1999vivo,Dombeck2010,Chen2013}.  
Calcium imaging has several important advantages: it offers high spatial resolution and can be coupled with various genetic tools to achieve cell-type specificity; it has proven to be scalable and can monitor hundreds/thousands of neurons in vivo simultaneously; finally, it allows for longitudinal tracking of cellular activity across multiple days or weeks~\cite{Ziv2013,Rubin2015,Driscoll2017}.  

At the same time, calcium imaging presents some important analysis challenges: calcium signals represent a slow, nonlinear encoding of the underlying spike train signals of interest, and therefore it is necessary to denoise and temporally deconvolve temporal traces extracted from calcium video data (\emph{e.g.}, $\Delta$ F/F) into estimates of neural activity.  These issues have received extensive attention in the literature \cite{Vogelstein2009,Vogelstein2010,Pnevmatikakis2016,Deneux2016,Theis2016,Friedrich2017,Jewell2018,Speiser2017,Aitchison2017,Berens2018,Pachitariu2018,Greenberg2018}.
Some of these deconvolution approaches estimate spiking probabilities directly \cite{Vogelstein2009,Pnevmatikakis2016,Deneux2016,Speiser2017,Aitchison2017,Greenberg2018}, but many approaches instead estimate the influx of calcium in each time bin, rather than a spiking probability \cite{Vogelstein2010,Pnevmatikakis2016,Friedrich2017,Jewell2018,Berens2018,Pachitariu2018,Stringer2019}; 
these non-probabilistic approaches tend to be faster and are therefore popular in practice. 

What is a proper statistical model for the output of these non-probabilistic calcium deconvolution approaches?  Defining such a model is the first step in any likelihood-based downstream analyses, \emph{e.g.}, Bayesian decoding, probabilistic latent factor modeling, and/or estimation of neural encoding models \cite{paninski2018neural}
--- but somewhat surprisingly, no ``standard model'' has emerged yet for the deconvolved output. 

The simplest approach is to simply threshold the output and treat the resulting super-threshold events as ``spikes'' (corresponding to a Bernoulli statistical model for these spikes), but this approach clearly discards information about the number of spikes per bin, and there is no obvious optimal way to set the threshold.  Another naive approach would be to apply standard point-process models (\emph{e.g.}, Poisson regression models) to the deconvolved output --- but as we will see below, the Poisson model is a poor approximation here, not least because the deconvolved output can take continuous values, while the Poisson distribution is supported on the integers.

In this paper, we investigate statistical models to characterize the deconvolved calcium activity.  (To be clear, we do not propose any new deconvolution approaches here; instead, we restrict our attention to modeling the output of existing deconvolution methods.)  In particular, we propose a zero-inflated gamma (ZIG) model, a two-component mixture model including a ``spike'' of probability at zero response and another continuous component for modeling positive responses, specified by a gamma distribution.  We apply this model to simulated data and real imaging datasets from hippocampus and thalamus, and find that it provides good fits across a wide variety of deconvolution parameters and data types.  Next we show that the ZIG model can be embedded within ``encoding models'' to characterize the probability of calcium responses given time-varying covariates such as the location or orientation of the animal during behavior.  Finally, we demonstrate that the ZIG-based encoding model leads to more accurate Bayesian decoding of these covariates. 

\section{Results}

\begin{figure}[ht!]
\centering
\includegraphics[width=0.8\linewidth,clip=true]{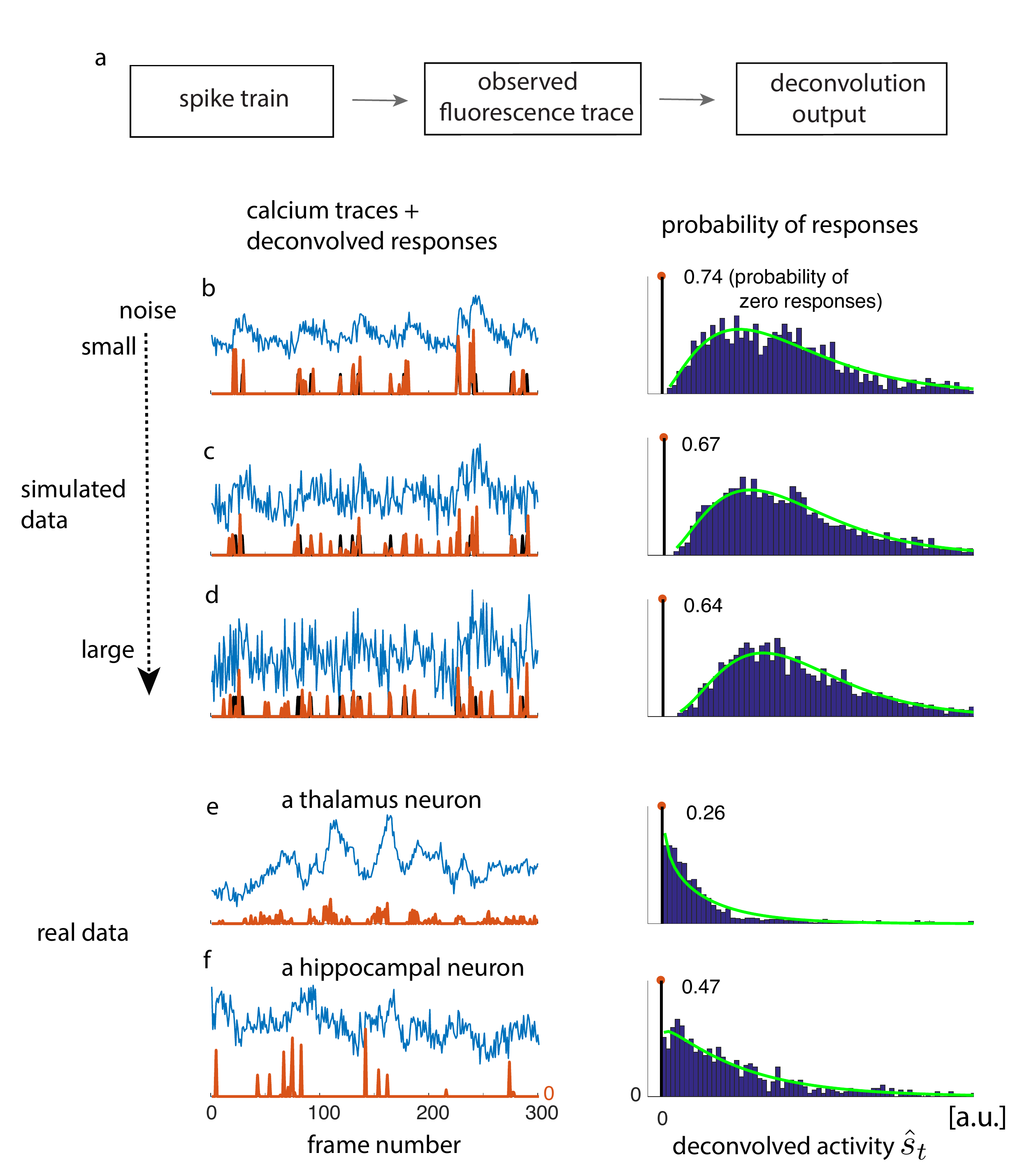}
\caption{{Illustration of the zero-inflated gamma (ZIG) model: deconvolved calcium responses typically consist of a mixture of zero responses plus a continuous component that is well-modeled as a gamma distribution, in both simulated and real imaging data.
{(a)} Pipeline of the simulations for the generation of artificial data. The spike train is sampled from a inhomogeneous Poisson process. The calcium concentration is then determined by a auto-regressive process (AR($1$) process, with decay time constant $450$ms), driven by the spike train.  The observed calcium trace is determined by the calcium concentration plus independent Gaussian noise. The deconvolved calcium responses are obtained using the OASIS deconvolution algorithm  described in \cite{Friedrich2017}.
(b,c,d) Left: observed fluorescence trace (blue), ground truth spikes (black), and the deconvolved output (orange). Each frame = $30$ms.   {Right}: the histogram of the deconvolved output (blue) and the ZIG fit (green); the number on each histogram represents the proportion of zero responses. The additive noise level of the simulated fluorescence increases from panel b to panel d. (e) Observed fluorescence and deconvolved response of a neuron from ADN (see Section~\ref{sec:method} for full experimental details). (f) Same as panel e but from the hippocampus (again, see Section~\ref{sec:method} for full experimental details). (Conventions as in b-d but we no longer have access to the true spikes.)  In each case, the ZIG model provides a good fit to the deconvolved outputs. 
}}
\label{fig:noise}
\end{figure}

\subsection{Nonnegative deconvolution methods applied to calcium fluorescence traces produce a mixture of zeros and positive real-valued output, well-captured by the zero-inflated gamma model}

\begin{figure}[ht!]
\centering
\includegraphics[width=1\linewidth,clip=true]{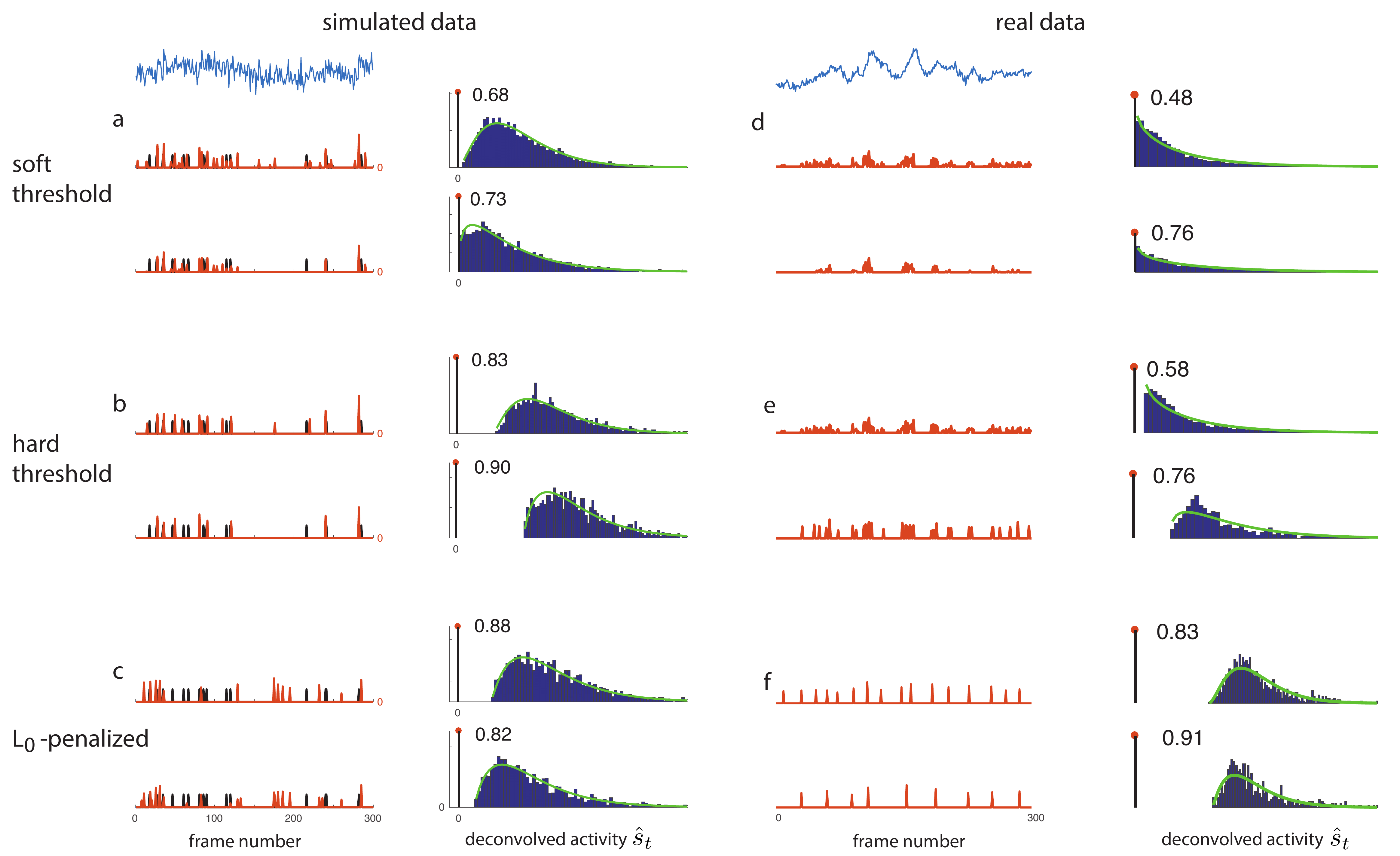}
\caption{The ZIG model is robust with respect to the details of different deconvolution methods. We consider three deconvolution methods, including a $L_1$-penalization method with soft threshold (panel a,d) \cite{Pnevmatikakis2016}, a method with non-zero minimal spike size or hard threshold (panel b,e) \cite{Friedrich2017}, and an $L_0$-penalized method (panel c,f) \cite{Jewell2018}. We apply these methods to both simulated  (panel a,b,c) and real data (panel d,e,f; same traces as in Figure \ref{fig:noise}). Each method has a hyper-parameter controlling the sparseness of the deconvolved activity. Two values of the sparseness parameter are examined for each method. Conventions as in Figure \ref{fig:noise}.}
\label{fig:inferencemethod}
\end{figure}

We begin with simple simulated data (Figure~\ref{fig:noise}a-d): we generate a Poisson spike train, then push this spike train through a standard auto-regressive AR($1$) model for calcium response \cite{Vogelstein2010} and add noise to generate simulated fluorescence traces, and then run a popular non-negative deconvolution method \cite{Vogelstein2009,Friedrich2017}
to obtain the post-deconvolution response, denoted as $\hat{s}_{t}$.  With experimentally relevant signal-to-noise levels, the resulting histogram of deconvolved responses $\hat{s}_{t}$ typically has a ``spike-and-slab'' form (Figure~\ref{fig:noise}b-d, right): significant mass is placed exactly at zero (the ``spike''), with the remaining mass forming a continuous ``slab'' on the positive real axis.  (This spike-and-slab structure of $\hat{s}_{t}$ is unsurprising: the deconvolution approach applied here enforces sparsity and non-negativity constraints on $\hat{s}_{t}$, forcing $\hat{s}_{t}$ to be exactly zero for many timesteps $t$.)  Empirically, a shifted gamma model suffices to capture the shape of the slab (green traces in Figure~\ref{fig:noise}b-d, right); the shift is fixed to be equal to the minimum spike size allowed by the deconvolution algorithm (``$s_{min}$''), and therefore the gamma distribution is still specified by two parameters.  We denote the resulting three-parameter distribution (with the third parameter corresponding to the probability of a non-zero response) as the ``zero-inflated gamma'' (ZIG) model. As we will see below, it is critical to use at least a two-parameter distributional family for the slab, to capture changes in the mean and variance.  The gamma family is a convenient two-parameter family that provides a good fit to the data, but other distributional families beyond the gamma could also be suitable here.

In these simulations, we have made several simplifying assumptions, including: i) a simple AR($1$) model for the generative process of the calcium fluorescence; ii) the a priori knowledge about the time constant of the AR process; iii) the increase of calcium concentration following each spike has a constant, deterministic size.
Presumably, any deviations from these assumptions would make the estimated ``spikes'' $\hat{s}_{t}$ noisier, thus making the continuous part of the response histogram smoother.  On the other hand, increasing the signal-to-noise ratio (SNR) and/or making the neurons burstier can introduce multiple ``bumps'' in the continuous part of the distribution (not shown).  This multiple-bump case could potentially be handled by incorporating a multiple-component mixture model for the slab in our spike-and-slab model, but (as we will discuss next), in practice for real data we have not found this to be necessary and have not pursued this direction systematically. 

We turn next to real data.  In most real datasets, the ground truth spiking (the first part of our simulation pipeline outlined in Figure \ref{fig:noise}a) is not available, but nonetheless we can run the same deconvolution algorithm on the observed fluorescence trace to obtain $\hat{s}_{t}$.  The resulting histogram of $\hat{s}_{t}$ is again well-fit by the ZIG model, for two example neuron cases shown in Figure \ref{fig:noise}e-f.

\subsection{The ZIG model is applicable to the outputs of multiple deconvolution methods, applied to data from multiple calcium indicators}

We next seek to determine whether the observations made in Figure \ref{fig:noise} are specific to a particular deconvolution method or calcium indicator. 

\begin{figure}[ht!]
    \centering
    \includegraphics[width=1\textwidth, clip=true]{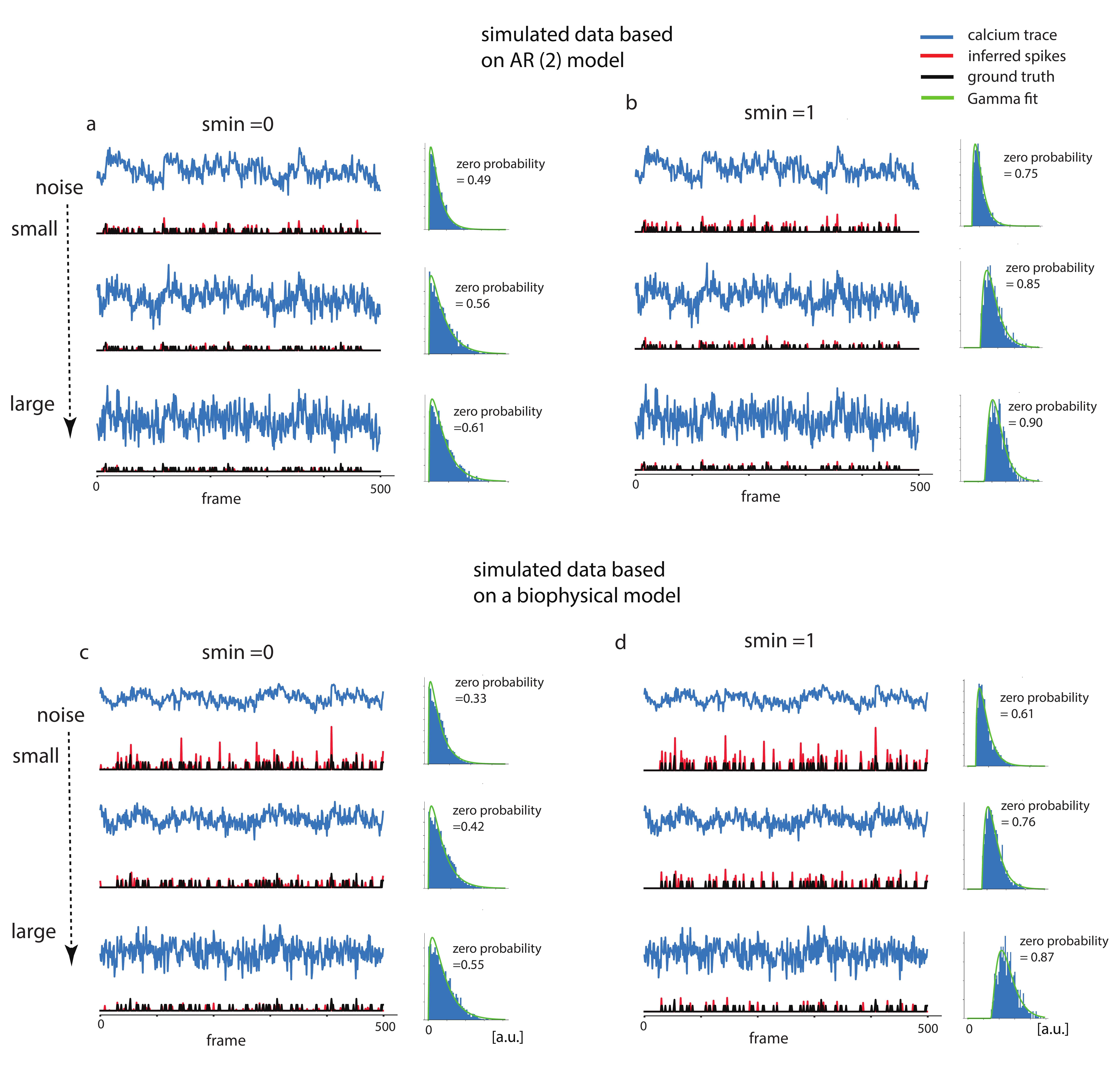}
    \caption{The ZIG model provides a good fit to the deconvolved calcium responses of simulated data based on a AR($2$) model and a more biophysical realistic model. The spike train is sampled from a homogeneous Poisson process.  (a,b) Results based on an AR($2$) model under three different noise levels and two different smin values in deconvolution~\cite{Vogelstein2010,Pnevmatikakis2016,Friedrich2017}. Left: fluorescence trace from simulations (blue), ground truth spikes(black), and the deconvolved output (red). Right:  the histogram of the positive deconvolved output (blue) and the ZIG fit (green). (c,d) Similar to (a,b), but based on a more biophysical realistic model as described in~\cite{Lutcke2013}.
    }
    \label{fig:SI_generative}

\end{figure}

In Figure \ref{fig:inferencemethod}, we examine three deconvolution methods, including an $L_1$-penalized method with a soft threshold \cite{Vogelstein2010,Pnevmatikakis2016}, a method with a hard threshold (i.e., positive minimal spike size ``$s_{min}$'') \cite{Friedrich2017}, and an $L_0$-penalized method \cite{Jewell2018a,Jewell2018}. Each method has a free parameter which controls the sparsity of the inferred responses; varying this parameter leads to corresponding changes in the histograms of the deconvolved responses $\hat{s}_{t}$, with more or less probability mass assigned to $\hat{s}_{t}=0$.  Over a range of parameters, the ZIG model provides a good fit to the output histogram for all three of the algorithms examined here. We also found that the ZIG model provides a good fit to the output of deconvolution applied to data generated from an AR($2$) model as well as the more biopysically detailed model from \cite{Lutcke2013} (see SI Figure~\ref{fig:SI_generative}).

\begin{figure}[ht!]
\centering
\includegraphics[width=1\linewidth,clip=true]{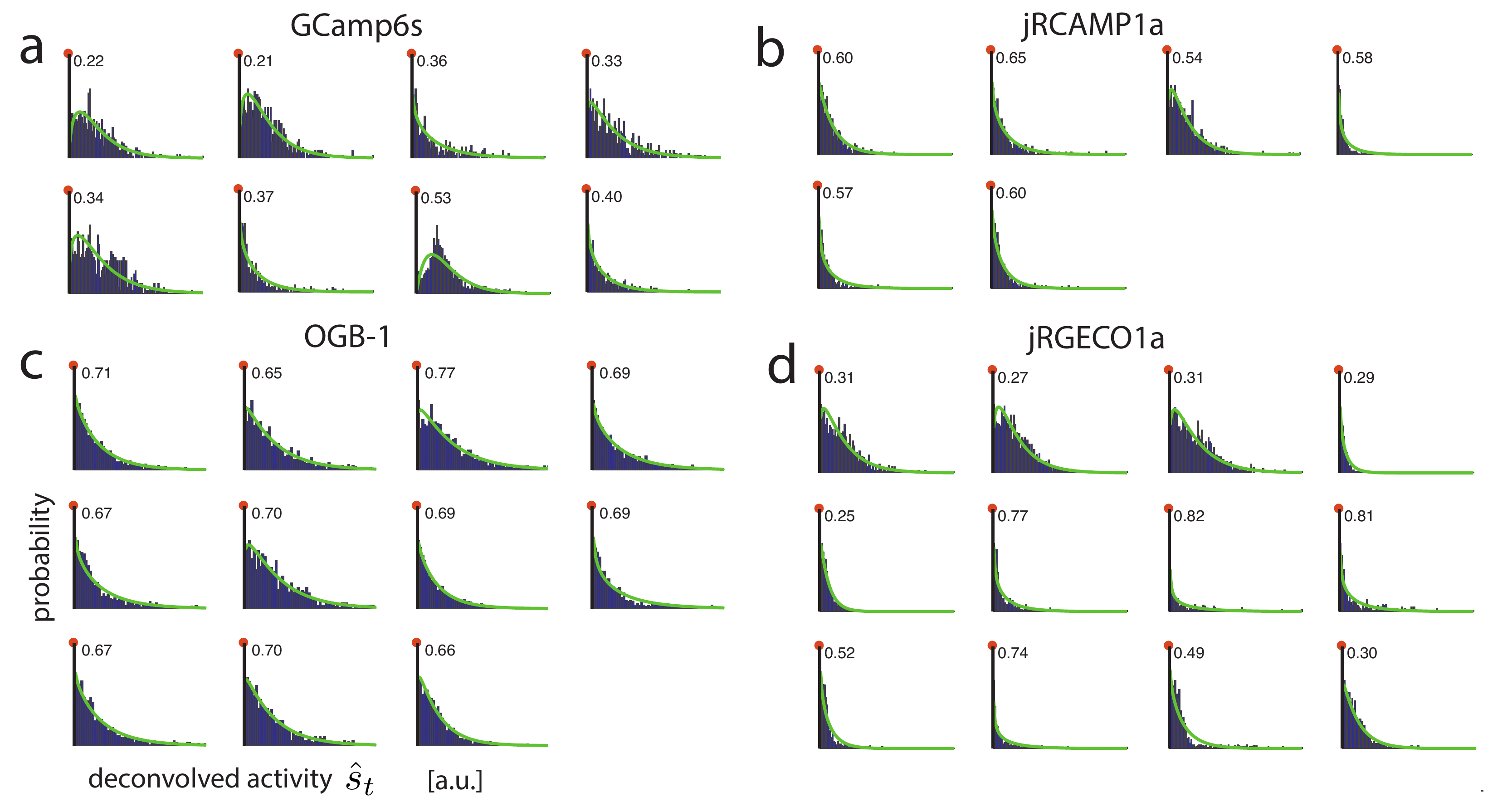}
\caption{The ZIG model is robust with respect to data collected with different calcium indicators. Data collected using four different calcium indicators are tested. The data are from the SpikeFinder challenge dataset (panel a,c from \cite{Theis2016}; panel b,d from \cite{Dana2016}). (Specifically, the four datasets used here are SpikeFinder dataset \#1 (calcium indicator OGB-1), dataset \#3 and \#5 (indicator GCamp6s), dataset \#9 (indicator jRCAMP1a), and dataset \#10 (indicator jRGECO1a).) All the neurons examined here fired at least 200 spikes.}
\label{fig:indicator}
\end{figure}

Next, in Figure \ref{fig:indicator}, we examine data shared through the SpikeFinder challenge \cite{Berens2018}, including traces recorded using four calcium indicators (GCamp6s, jRCAMP1a, OGB-1, jRGECO1a).  Again, we find that the ZIG model provided a good fit across a wide range of data.

\begin{figure}[ht!]
\centering
\includegraphics[width=1\linewidth,clip=true]{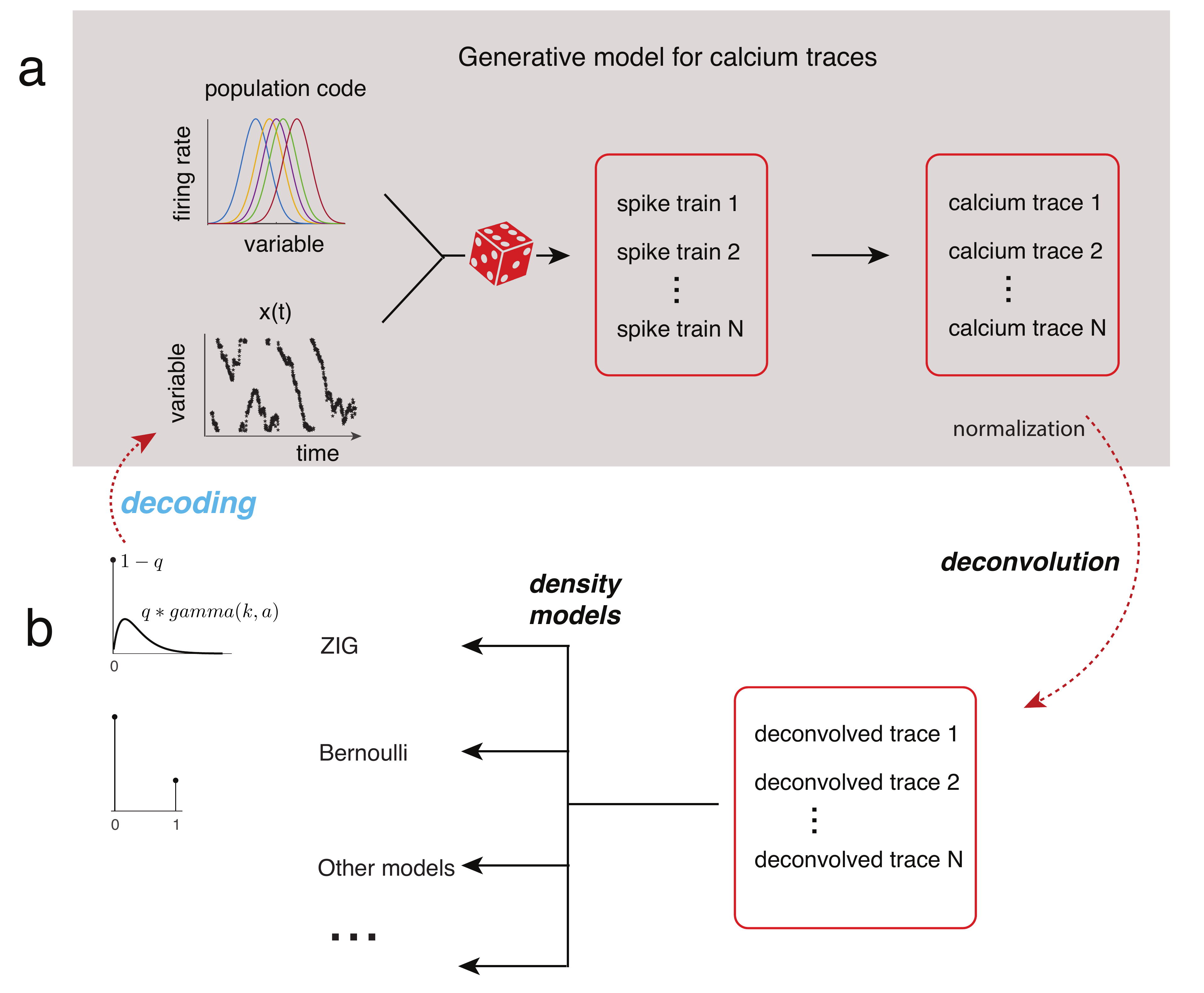}
\caption{The encoding-decoding modeling framework.  (a) The full generative ``encoding'' model for calcium fluorescence traces.  Given the neural tuning curves and the covariate sequence $\theta_t$, spikes are generated probabilistically and transformed (as in Figure \ref{fig:noise}) into the observed fluorescence traces. 
(b) After deconvolving to obtain $\hat{s}_{t}$, we can use the ZIG as well other models for $\hat{s}_{t}$, and use the estimated encoding model and $\hat{s}_{t}$ to decode $\theta_t$. 
}
\label{fig:overview}
\end{figure}

\subsection{Constructing encoding models for simulated calcium responses}

We have seen above that 
the ZIG model provides a good fit to the marginal distribution of the deconvolved responses $\hat{s}_{t}$.
Now we want to exploit this probabilistic model to perform neural data analysis tasks.
The first step is to 
fit \emph{encoding models}: ie, what is $p(\hat{s}_{t} | \theta_{t})$, for some observed covariate $\theta_{t}$ such as a stimulus or movement. In general, $\theta_t$ may be multi-dimensional, but in the example applications here $\theta_t$ will be one dimensional. 
Once these encoding models are fit and validated, we can use them to perform tasks like decoding of $\theta_{t}$ given the observed deconvolved responses $\hat{s}_{t}$.
The overall approach is illustrated in Figure \ref{fig:overview}.
 
To fit the ZIG model to $p(\hat{s}_{t} | \theta_{t})$, we need to fit three parameters, each of which may depend on $\theta$: the probability of non-zero response $q(\theta)$, the scale parameter $a(\theta)$, and the shape parameter $k(\theta)$ for the gamma component, specifically  
\begin{equation}
p\left(\hat{s}_{t} | \theta_{t}\right) = \left(1-q(\theta_{t})\right)\cdot\delta(0) + q(\theta_{t})\cdot\textrm{gamma} \left(\hat{s}_{t}; k(\theta_{t}), a(\theta_{t}), loc\right),
\end{equation}
where again we fix the location parameter $loc$ for the gamma component as the minimum spike size $s_{min}$. (The mean of the ZIG model with parameters $(q,k,a, loc)$ is $q(ka+loc)$, and the variance is $qka^2+q(ka+loc)^2(1-q)$.) We model these parameters as nonlinear functions of $\theta_{t}$; we use neural networks to parameterize these nonlinearities, and then estimate the weights of these networks by maximum likelihood (see Section~\ref{sec:method} for full details).  

\begin{figure}[ht!]
    \centering
    \includegraphics[width=1\textwidth, clip=true]{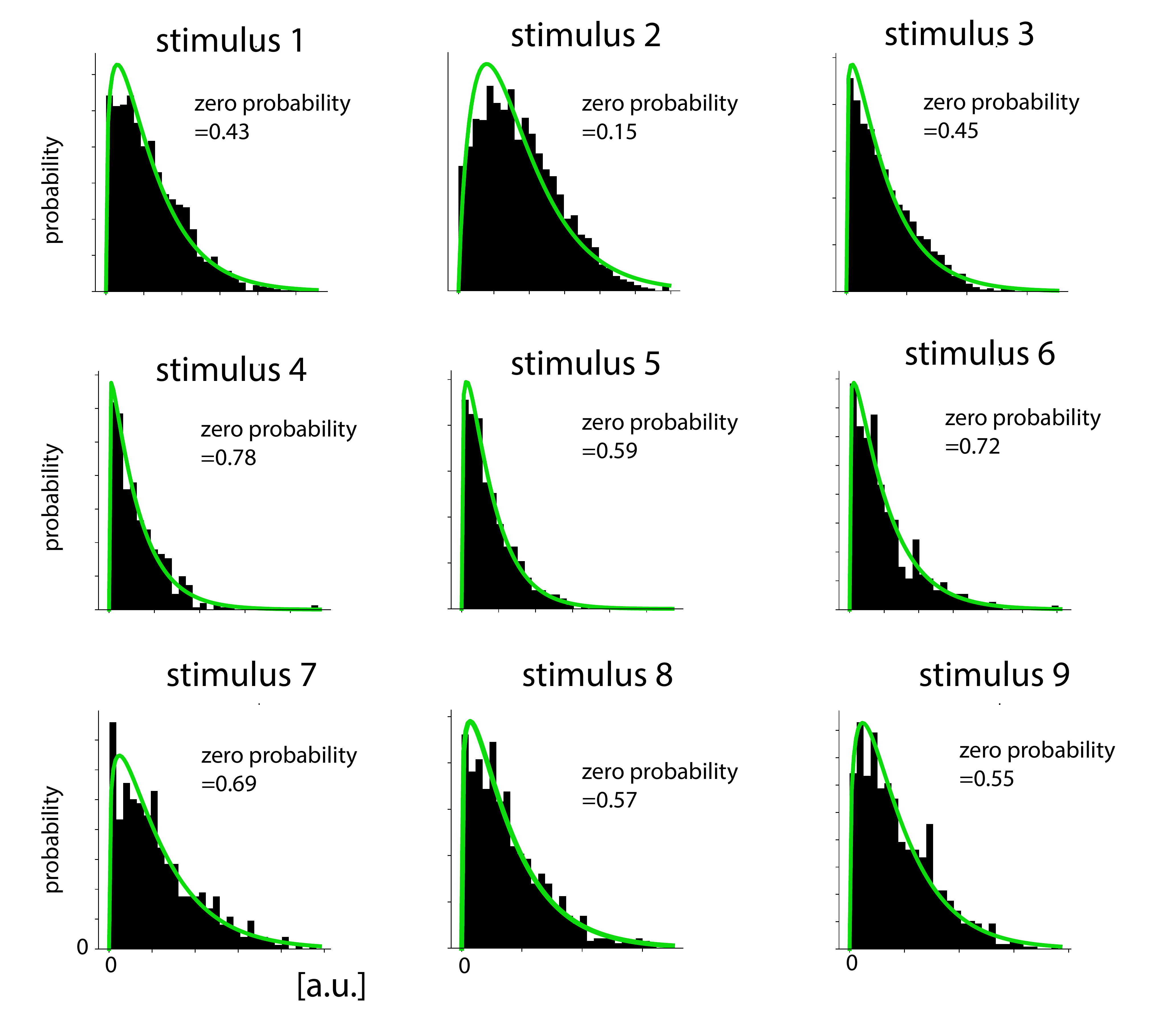}
    \caption{The ZIG model provides a good fit to the conditional probability distribution of deconvolved calcium responses given stimulus. The spike train is sampled from an inhomogeneous Poisson process with firing rate modulated by the stimulus, and the calcium concentration is determined by an AR($1$) model. Each panel shows the histogram of the positive deconvolved output corresponds to each stimulus value. The green line is the ZIG fit.}    \label{fig:SI_local}

\end{figure}

It is worth pausing to note two points here.  First, the general problem of estimating a mixture model whose parameters depend on $\theta_t$ would be rather challenging; however, in our case we are fitting a very particular two-component mixture model in which the first component (corresponding to $\delta(0)$) is trivial to estimate, since we merely need to count zero values in $\hat s_t$, resulting in a much easier estimation problem.  Second, in the previous sections we showed that the ZIG provides a good \emph{marginal} fit to $p(\hat s_t)$, marginalized over the whole dataset --- but there is no guarantee that the same model provides a good \emph{conditional} fit to $p(\hat s_t | \theta_t)$.  We have checked this fit empirically, and it turns out that the ZIG model also provides a good fit to the conditional distributions considered here, in both real and simulated data (see SI Figure~\ref{fig:SI_local} for an example).
   
To test this conditional estimation approach, we generate artificial calcium imaging datasets with hundreds of simulated neurons (Figure \ref{fig:overview}). We first construct tuning curves of individual neurons that tile the space of $\theta$ values.  In the real data examples presented below, $\theta$ will be a one-dimensional variable (\emph{e.g.}, the animal's head direction), so we use a one-dimensional $\theta$ in these simulations. Next we take a empirically measured time series $\theta_{t}$ (the head direction of a mouse), and compute the time-varying firing rates for individual neurons by plugging $\theta_{t}$ into the tuning curves. We then generate binned spike trains with different noise characteristics; we experiment with spike counts drawn from a Poisson distribution or a negative binomial (NB) distribution, as both have been proposed to model empirically observed spike responses \cite{Tomko1974,Tolhurst1981,Goris2014}.  Next we plug these simulated binned spike trains into the same generative model for calcium fluorescence traces discussed above, then deconvolve the resulting traces to obtain simulated responses $\hat{s}_t$.
Finally, we fit the ZIG encoding model to the resulting  responses $\hat{s}_t$.

\begin{figure}[ht!]
    \centering
    \includegraphics[width=1\textwidth,clip=true]{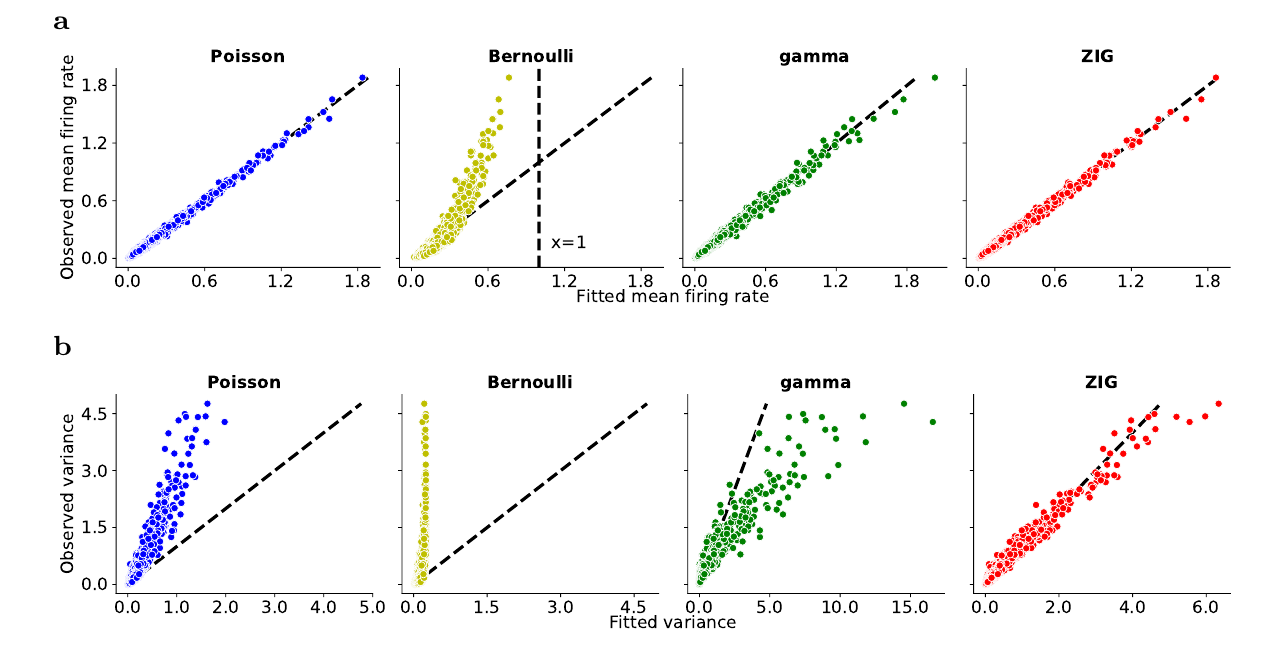}
    \caption{The ZIG model captures the means and variances of deconvolved calcium responses in the simulated data.
    (a,b) Summary plots based on all the neurons ($N=215$) showing the observed versus the predicted mean/variance of $\hat{s}_t$.  We divide the $\theta$ range to into small equi-spaced bins (number of bins $= 18$ here), and compute the mean and variance for the observed and fitted responses corresponding to each bin for each model neuron. Each dot represents the mean (or variance) associated with one bin from one neuron.}
    \label{fig:sim_nb_enc}
\end{figure}

We compare the ZIG model against simpler Poisson, Bernoulli, and gamma models (see Section~\ref{sec:method} for full details). Figure \ref{fig:sim_nb_enc} shows the results from a simulated dataset with negative binomial spiking. (We find that the results on the Poisson dataset are  qualitatively similar; data not shown.)
Overall, for both the Poisson or NB simulated datasets, we find that all of these models except for the Bernoulli model can capture the data mean well (the Bernoulli model is only effective for data in which the mean of $\hat{s}_t$ in each bin is bounded below $1$; this model fails to capture the responses in bins with high firing rates). However, only the ZIG model can properly capture both the mean and variability of $\hat{s}_{t}$.  (The ZIG model also provides a good empirical fit to the full conditional distribution; data not shown.)  The alternative models tend to either over- or under-estimate the variance, therefore providing poor descriptions of the distributions of the deconvolved responses; thus, the extra flexibility (due to the larger number of parameters) in the ZIG model is necessary to capture basic statistics of the data.  

\subsection{The ZIG encoding model leads to improved Bayesian decoding in simulated data}

In the previous section we showed that the ZIG encoding model is flexible enough to capture the mean and variance of $\hat{s}_{t}$ across a wide range of firing rate regimes, in simulated data.  Can we exploit this encoding model to obtain an improved decoder for $\theta_{t}$?  We use a classic Bayesian decoding  approach to address this question: we compute the posterior distribution of $\theta_{t}$, under the different encoding models for $\hat{s}_{t}$ discussed above, and then quantify how well the resulting posterior distributions capture the uncertainty in $\theta_{t}$ given the observed $\hat{s}_{t}$.  

\begin{figure}[ht!]
    \centering
    \includegraphics[width=1\textwidth, clip=true]{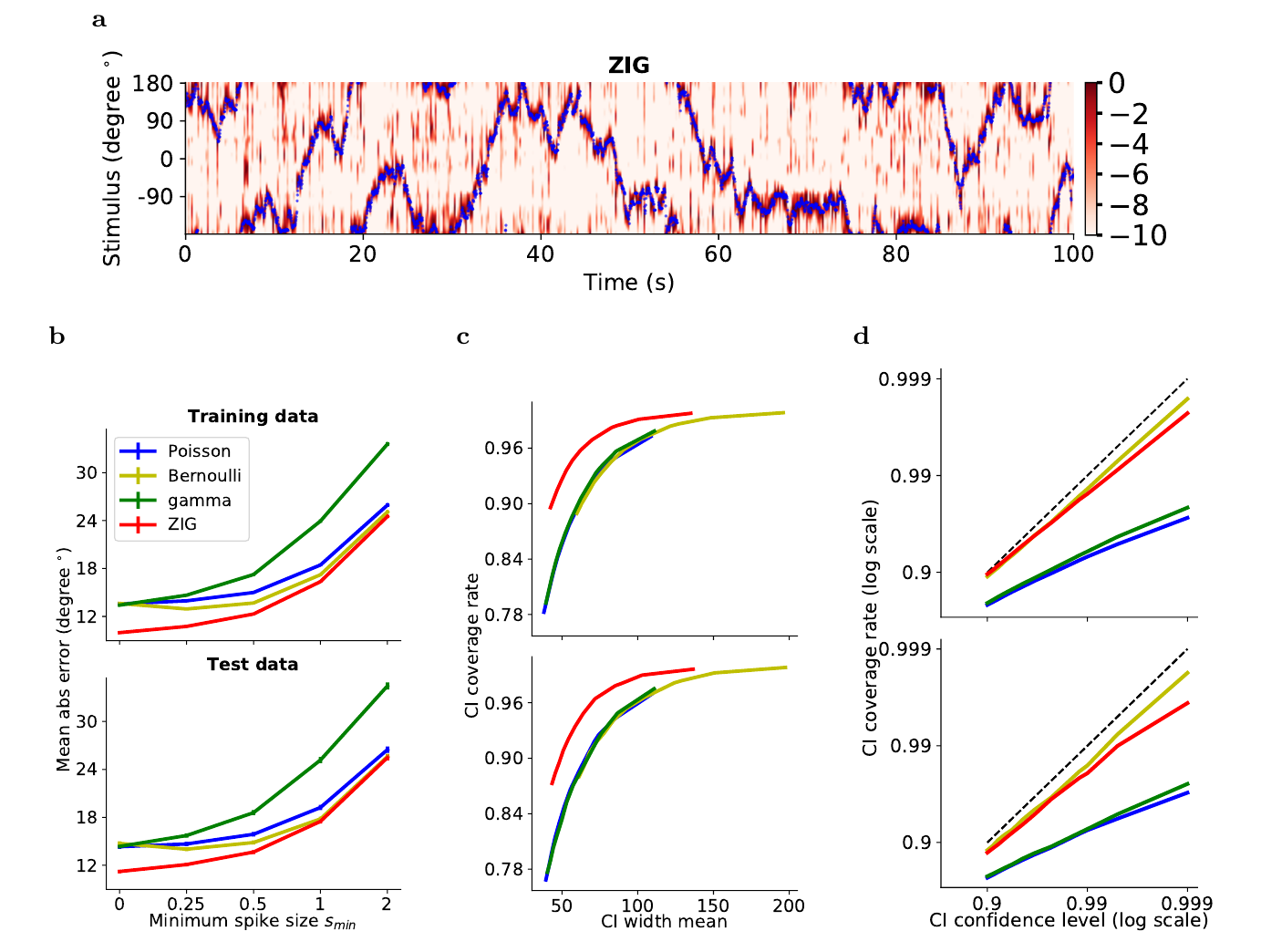}
    \caption{The ZIG model leads to improved decoding performance on simulated data. Decoding is performed based on the deconvolved responses in single frame, time window $\sim 33$ms, with no smoothing across frames on all neurons ($N=215$).
    (a) True simulated location (blue) plotted with the decoded normalized log-posterior probability under the ZIG model (red; posterior at each step is normalized to have a maximum of 1, for easier visualization).  Note that the decoded posterior does a good job of tracking the true location. (b) Decoding mean absolute error $\pm 1$ standard error under different encoding models, with varying $s_{min}$, the minimum spike size parameter in \cite{Friedrich2017}; larger values of $s_{min}$ correspond to sparser output $\hat{s}_{t}$.   (c) Posterior credible interval (CI) width vs CI coverage rate (the probability that the true location falls within the CI; higher is better here). (d) Confidence level vs CI coverage rate.  Dashed line indicates unity (i.e., the CI is achieving its nominal coverage rate).  In (b,c) we see that the ZIG model leads to the lowest decoding error and the best coverage rate over a range of parameters, while (d) shows that the CI computed under the ZIG model achieves a nearly-nominal coverage rate, as desired; in contrast, the Poisson and gamma encoding models output mis-calibrated credible intervals.
    }
    \label{fig:sim_nb_dec}
\end{figure}

We quantify the performance of the resulting decoders on simulated data in Figure \ref{fig:sim_nb_dec}.
Overall, the ZIG model leads to the smallest decoding error over a wide range of deconvolution sparsity parameters.  Interestingly, accuracy degrades monotonically as a function of the sparsity of the output $\hat{s}_{t}$: i.e., the decoders can take advantage of even very small outputs $\hat{s}_{t}$ to improve the decoding accuracy. (In Figure \ref{fig:sim_nb_dec} we use the deconvolution approach from \cite{Friedrich2017}, with a hard-threshold on the minimal spike size; results based on the soft-threshold deconvolution approach from \cite{Pnevmatikakis2016}  are similar.) The decoder based on the ZIG encoding model also achieves the highest coverage rate (i.e., the posterior credible interval covers the true value of $\theta_{t}$ with highest probability).  In contrast, the decoders based on the Poisson and gamma models output credible intervals with mis-calibrated coverage rates (i.e., the credible interval based on these models was narrower than it should have been), due to a mismatch between the true versus the modeled distribution of $\hat{s}_{t}$.  In other words, a Bayesian statistician using a Poisson or gamma encoding model would be (mistakenly) overly confident in her predictions.

\subsection{Application to real imaging
data} \label{sec:real}

\begin{figure}[ht!]
    \centering
    \includegraphics[width=1\textwidth,clip=true]{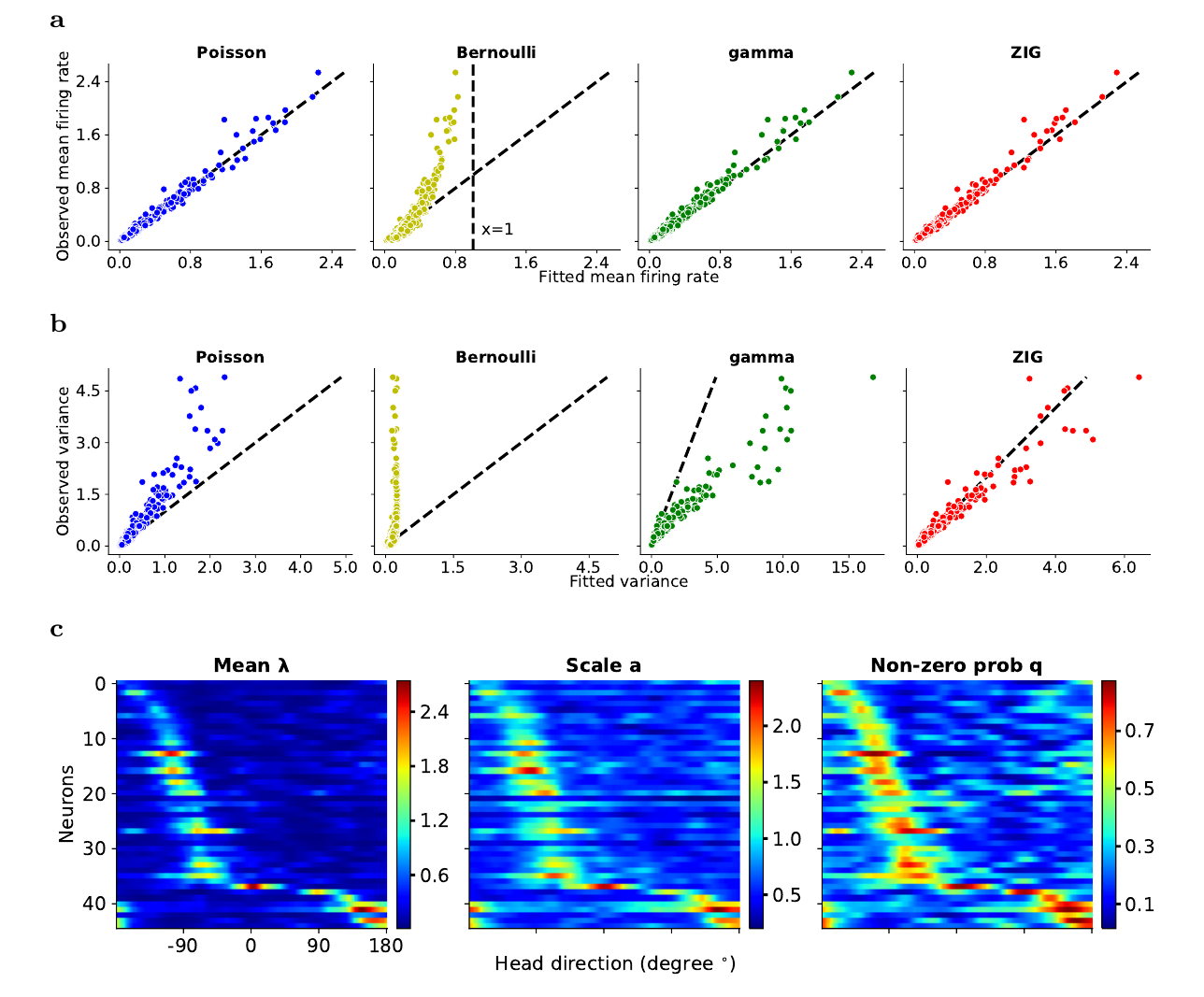}
    \caption{The ZIG model best captures the means and variances of deconvolved calcium responses in the ADN dataset. 
    (a,b) Conventions as in Figure~\ref{fig:sim_nb_enc}; similar to results based on the simulated data, the ZIG model again provides the best fits to the observed means and variances for this experimental dataset. (c) Estimated parameter values for the ZIG model and Poisson model, for multiple neurons (each row corresponds to one neuron, while the columns correspond to different $\theta$ values). Neurons are sorted according to the preferred firing direction. For the Poisson model, each row plots the mean firing rate $\lambda$ as a function of head direction for each neuron. For the ZIG model, there are three sets of parameters. The scale parameter $a$ and the probability of non-zero response $q$ are plotted. Notice that the two parameters are correlated; both parameters tend to scale with the estimated mean rate $\lambda$. We find that we could obtain good fits by fixing the shape parameter $k$ in the ZIG model for each neuron.}
    \label{fig:real_hd_enc}
\end{figure}

In the previous section we developed the encoding-decoding analysis pipeline on simulated data. Next we apply these methods to real data.  We focus on two calcium imaging datasets in this section.  The first is a single-photon imaging dataset collected from thalamic region ADN, and the second is a two-photon dataset from hippocampal region CA1.  Both datasets are collected in animals performing spatial navigation tasks (see Section~\ref{sec:method} for full details).  Our aim is to decode head direction (during free behavior) in the ADN data and location along a circular track (during head-fixed behavior) in the hippocampal data; thus in both cases the variable $\theta$ is one-dimensional, as in the simulated data.

\begin{figure}[ht!]
    \centering
    \includegraphics[width=1\textwidth,clip=true]{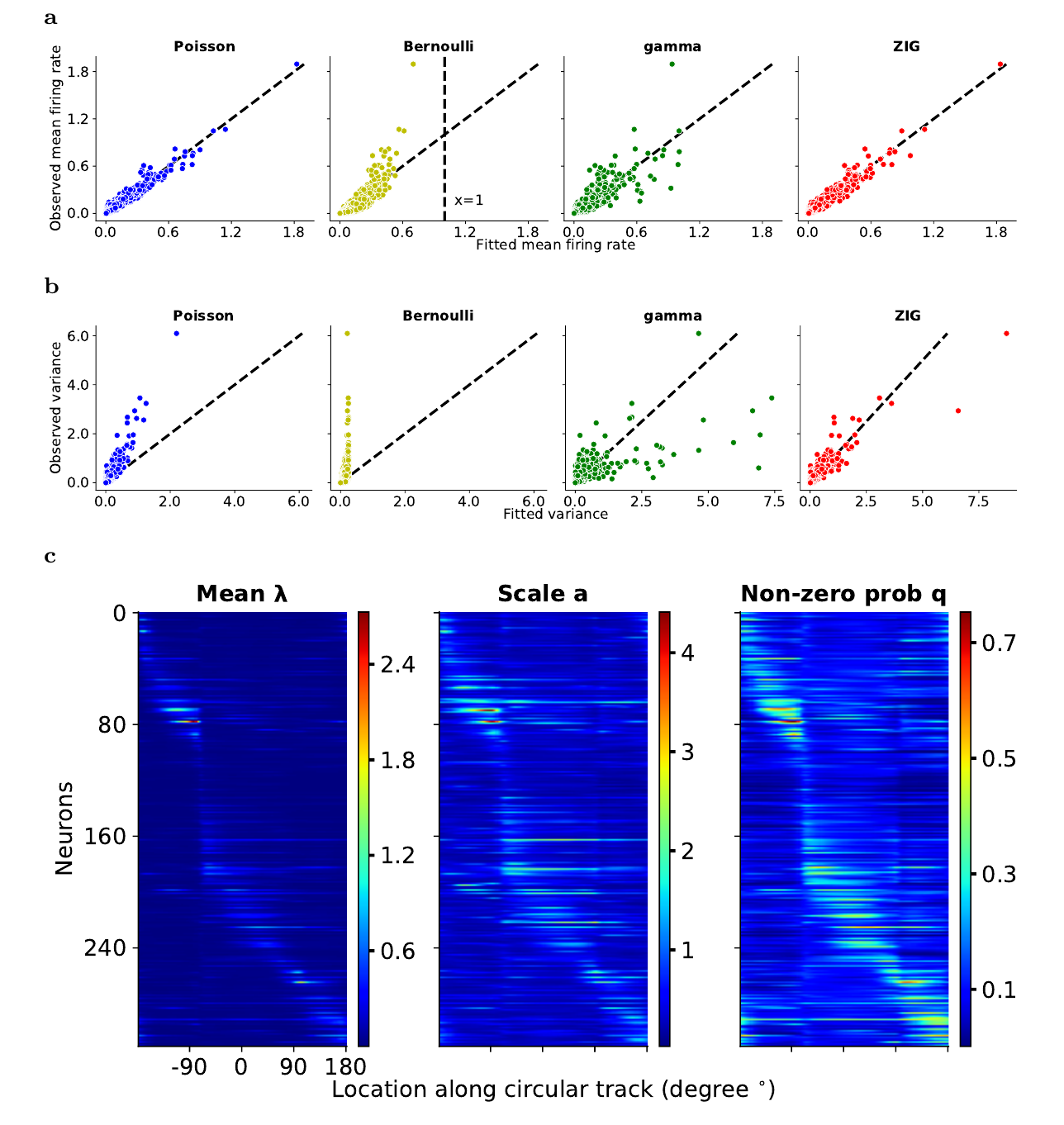}
    \caption{Encoding results for hippocampus data during running. Conventions as in Figure~\ref{fig:real_hd_enc}. (a,b) The observed versus the predicted mean/variance of $\hat{s}_t$. (c) Estimated parameter values for the ZIG model and Poisson model, for multiple neurons.  Again, among the four models considered, only the ZIG model can capture both the mean and variance of the calcium responses conditional on the animal's location on the track.}
    \label{fig:real_pos_enc}
\end{figure}

\begin{figure}[ht!]
    \centering
    \includegraphics[width=1\textwidth, clip=true]{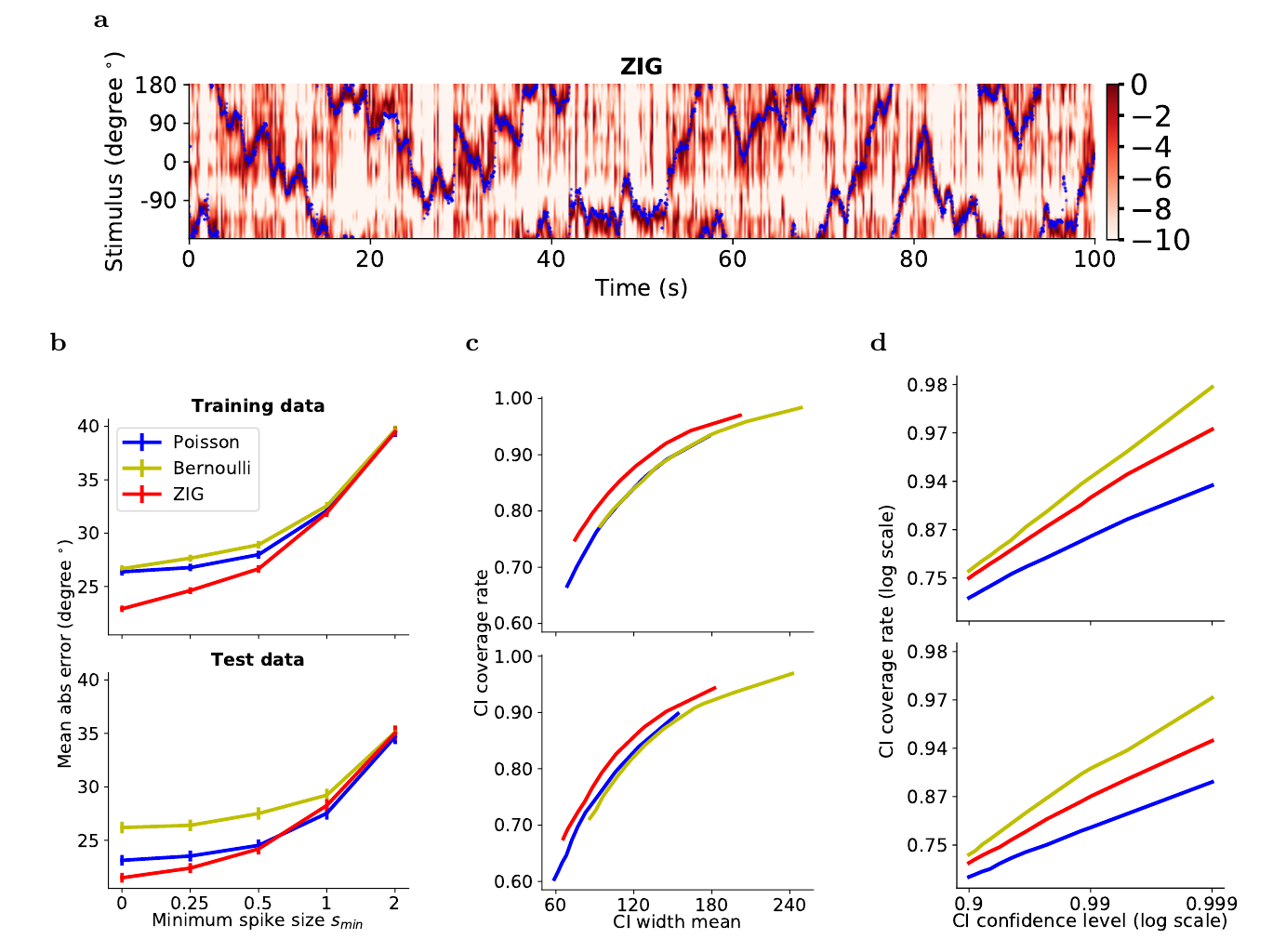}
    \caption{Decoding results for head direction (ADN) dataset. Decoding is performed based on responses from single frame, time window $\sim 33$ms, with no smoothing across frames on $N=45$ neurons. Conventions similar to Figure \ref{fig:sim_nb_dec}. (a) True simulated location (blue) plotted with the decoded normalized log-posterior probability for the ZIG model (red).  (b) Decoding mean absolute error $\pm 1$ standard error under different encoding models, with varying $s_{min}$.  (c) Posterior CI width vs CI coverage rate. (d) Confidence level vs CI coverage rate.  The gamma model performed poorly here and is not shown.}
    \label{fig:real_hd_dec}
\end{figure}

\begin{figure}[ht!]
    \centering
    \includegraphics[width=1\textwidth, clip=true]{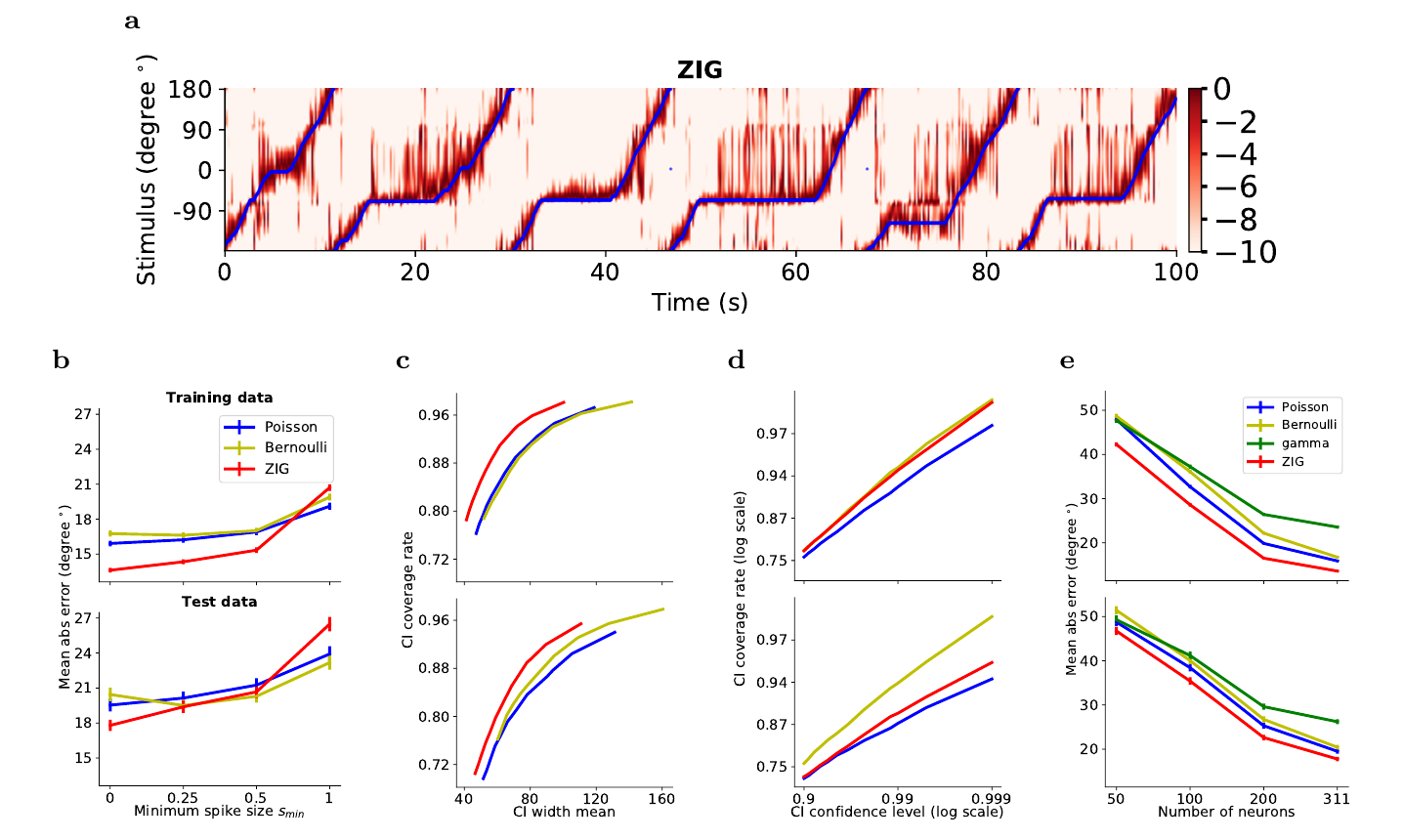}
    \caption{Decoding results for hippocampus data. The encoding model is estimated based on the data during running, as is typical in the hippocampal decoding literature \cite{Davidson2009}. Decoding time window is set to be every two frames, $\sim 33$ms. (a,b,c,d) similar to Figure~\ref{fig:real_hd_dec}. (e) Decoding results for subsets of neurons of hippocampus data. We randomly select neurons ($N=50,100,200$) out of the $311$ total neurons and perform the encoding-decoding analysis based on these subsets. The advantage of the ZIG model remains robust with smaller sub-populations of hippocampal neurons. }
    \label{fig:real_pos_dec}
\end{figure}

We begin in Figure \ref{fig:real_hd_enc} by fitting encoding models to the ADN data (see Figure~\ref{fig:real_pos_enc} for the results of hippocampal data).  The results are similar to those shown in  Figure \ref{fig:sim_nb_enc}: of the models examined here, only the ZIG model can capture both the mean and the variance of the empirical data.
Further, in panel c we examine the tuning curves from this population of neurons.  We compute the mean firing rates as a function of $\theta$ (leftmost panel) and plot these next to the estimated $a(\theta)$ and $q(\theta)$ curves (middle and right panels); recall that the mean of $\hat{s}$ as a function of $\theta$ scales proportionally with $a(\theta) q(\theta)$ in the ZIG model.  We see that the parameters $a(\theta)$ and $q(\theta)$ covary across this population, indicating that there may be some statistical benefit in fitting these parameters with a hierarchical model that can share information between $a(\theta)$ and $q(\theta)$; however, we have not pursued this direction systematically.

Next we turn to decoding (Figures \ref{fig:real_hd_dec} and \ref{fig:real_pos_dec}).  Again, the results
of the real data analysis are largely consistent with the simulated results presented in Figure \ref{fig:sim_nb_dec}: in both datasets, the ZIG encoding model leads to more accurate Bayesian decoding, with higher credible interval coverage rates.  Again, the decoding accuracy improves as the sparsity of $\hat{s}$ decreases.

One major difference between the simulated and real data is that the coverage probabilities are no longer well-calibrated, for any of the encoding models.  In other words, the Bayesian posterior based on these encoding models is overly confident.  We believe this is due to model mismatch: in our Bayesian decoder we model the responses $\hat{s}_{t}$ as conditionally independent across neurons and time given $\theta_t$, i.e., 
\begin{equation*}
    p \left( \{\hat{s}_{ti}\} | \theta_t \right) = \prod_{it} p \left( \hat{s}_{ti} | \theta_t \right), 
\end{equation*}  
where $\hat{s}_{ti}$ denotes the observed response at time $t$ from cell $i$.  This assumption makes a testable prediction: $\hat{s}_{ti}$ should be uncorrelated with $\hat{s}_{tj}$ (where $i$ and $j$ index two different neurons) if we restrict attention to responses within a single bin of $\theta$ values.  We find that these ``noise correlations'' are empirically not zero (invalidating the conditional independence assumption), and in fact if we perform a shuffling analysis in which we randomize the responses $\hat{s}_{ti}$ within each $\theta$ bin (thus preserving the relationship between $\hat{s}_{ti}$ and $\theta_t$ while destroying noise correlations between $\hat{s}_{ti}$ and $\hat{s}_{tj}$; see Section~\ref{sec:method} for full details), then we find that the calibration of the credible interval is restored (data not shown).  We leave further detailed modeling of these noise correlations to future work.

\section{Discussion}

The primary conclusion of this work is that the ZIG model provides a significantly improved fit to the distribution of the post-deconvolved calcium responses $\hat{s}_{t}$: the ZIG model is sufficiently flexible to capture the zero-inflation and varying mean and dispersion of the data across a wide variety of indicators, deconvolution methods, and behavioral settings.  Moreover, it is straightforward to extend this into a $\theta_{t}$-dependent encoding model, and in turn to use this encoding model for Bayesian decoding.  The improved encoding fits provided by the ZIG lead directly to more accurate decoding, with better-calibrated posterior uncertainties. Finally, somewhat surprisingly, we find that setting the deconvolution hyperparameter to minimize the sparsity of $\hat{s}_{t}$ consistently leads to the most accurate decoder (consistent with results in \cite{Pachitariu2018}); i.e., attempting to discard small ``noisy spikes'' in $\hat{s}_{t}$ may be counterproductive.
Overall, the ZIG model fills a crucial gap for calcium imaging analyses, by providing a firm statistical foundation for encoding and decoding models based on the estimated activity $\hat{s}_{t}$.  

Of course the two-step approach followed here --- to deconvolve the observed fluorescence traces, fit a probabilistic model to the deconvolved output, and then use this model to compute likelihoods in downstream encoding and decoding models --- is not the only option.  Another approach would be to use a full statistical model of the observed fluorescence traces (instead of treating the deconvolved output $\hat s_t$ as the observed data, as we did here), as in \emph{e.g.} \cite{Ganmor2016,Aitchison2017,Speiser2017,triplett2019model}.
This ``end-to-end'' modeling approach has the advantage that it can model more complex temporal dependencies in $\hat s_t$, and can potentially use side information to obtain better estimates of the neural activity from noisy fluorescence observations (see \cite{Vogelstein2009,Vogelstein2010,picardo2016population,mishchenko2011bayesian,Pnevmatikakis2016} for further examples along these lines).
Conversely, there are a number of cases for which deconvolution or sophisticated statistical modeling is not required at all to address the scientific question at hand.
The two-step approach pursued in this paper can be seen as a useful compromise between these two extremes: if the researcher's scientific question requires more temporal resolution than is available from the raw fluorescence measurements (i.e., deconvolution is necessary), but the researcher lacks the time or expertise needed to develop, estimate, and test a full end-to-end statistical model, then the two-step approach developed here offers a quick, effective, practical compromise.

Open source code implementing the methods presented here along with the sample datasets is available at \url{https://github.com/zhd96/zig}.
We hope these methods will be useful for the variety of downstream analyses that are currently being pursued by the calcium imaging community.

\section{Materials and Methods}\label{sec:method}

\subsection{Density models of the deconvolved calcium trace} 

\subparagraph{ZIG model} We model the density of the devonvolution output as
\begin{equation}
\hat{s}_{ti}\sim (1-q_{ti})\cdot\delta(0) + q_{ti}\cdot\textrm{gamma}(k_{ti},a_{ti},loc_{i}),
\end{equation}
where $q_{ti}$ denotes the probability of non-zeros, $a_{ti}$ is the scale parameter of the gamma distribution, and $k_{ti}$ is the shape parameter of the gamma distribution, for neuron $i$ and time $t$. $loc_{i}$ is the location parameter of the gamma distribution for neuron $i$, fixed as the minimum spike size $s_{min}$.  We denote $a_i=(a_{1i},\cdots,a_{Ti})^{\top}\in\mathbb{R}_{+}^{T}$ as the scale parameters for neuron $i$. Parameters $q$ and $k$ are defined similarly. We denote this family of density functions as $\hat{s}\sim \textrm{ZIG}(q,k,a)$.
Note that when $s_{min}=0$, the ZIG density family has a useful scale-invariance property: if $\hat{s}\sim \textrm{ZIG}(q,k,a)$, then $c\hat{s}\sim \textrm{ZIG}(q,k,ca)$, for any constant $c>0$. This is convenient because in general the scale factor connecting spikes to increases in calcium concentration is unknown (and will typically vary from cell to cell); however, the scale invariance of the ZIG model implies that we do not need to estimate this scale factor explicitly.

\subparagraph{(Scaled-)Poisson model} The Poisson model places all of its probability mass on the non-negative integers, and is therefore inappropriate for modeling $\hat{s}_t$, which has range $\mathbb{R}_{+}$.  Nonetheless, as discussed below, it is possible to assign a pseudolikelihood under the Poisson model to real-valued observations $\hat{s}_t$, and to fit the Poisson rate $\lambda$ by maximizing this pseudo-likelihood.  However, the Poisson model does not have the scale-invariance property enjoyed by the ZIG model, and therefore some care must be taken in defining a scale for $\hat{s}_t$ (empirically, we find that the performance of the Poisson encoding and decoding models are highly sensitive to scaling of $\hat{s}_t$).  We experiment with two scaling approaches. 
In the first scheme, the deconvolved trace is normalized by the noise standard deviation of the raw calcium trace, using methods proposed previously~\cite{Pnevmatikakis2016}.
In the second approach, the deconvolved trace $\hat{s}_t$ is normalized by its Fano factor.  Either of these normalizations leads to similar performance in terms of encoding or decoding accuracy (data not shown).

\subparagraph{Bernoulli model} 
The Bernoulli model can be considered as a special case of the ZIG model, by collapsing the positive responses into a delta function at $1$. 
The responses are first binarized by thresholding; as discussed in the main text, we explore a range of different $s_{min}$ values in the deconvolution step, and set the binarization threshold equal to $s_{min}$.  As in the ZIG model, we define $q_i=(q_{1i},\cdots,q_{Ti})^{\top}\in\mathbb{R}_{+}^{T}$ as the non-zero probability for neuron $i$.

\subparagraph{Gamma model} For completeness, we also fit a gamma model to the deconvolved responses.  (Note that the gamma distribution can not capture the strong bimodality that we typically observe in $\hat{s}_t$.)  The 
gamma distribution exhibits a singularity at 0 when the shape parameter $k$ is less than 1. To avoid this issue, we slightly shift the observations away from 0 by adding a small positive number $\epsilon$ ($\epsilon=10^{-4}$) to $\hat{s}$ before fitting the $\textrm{gamma}(k,a)$ model. As in the ZIG model, we denote $a_i=(a_{1i},\cdots,a_{Ti})^{\top}\in\mathbb{R}_{+}^{T}$ as the scale parameters for neuron $i$. 

\subparagraph{Parameter estimation} We estimate the parameters of the above models via maximum (pseudo-)likelihood. Details appear in Section~\ref{sec:model_fitting}.

\subsection{In vivo datasets}

Two in vivo datasets are analyzed here.  Both datasets are about 15 minutes long; in each case GCaMP6f was utilized as the calcium indicator.  Traces are extracted using the 
CNMF-E software described in \cite{zhou2018efficient}.

The first dataset is from area ADN of mouse thalamus. During stereotaxic surgery a male B6/C57j mouse was injected in ADN  with the viral vector AAV9-hSyn-GCaMP6f (Molecular Tools Platform, Laval University). These mice were then implanted with a GRIN relay lens that was 500 microns in diameter and 4.0 mm in length (Inscopix, Inc.). The lens was positioned such that the bottom surface of the lens terminated just dorsal to the ADN. Baseplates used to attach the miniaturized fluorescent imaging endoscope (`UCLA Miniscope', miniscope.org) were cemented to the skull and imaging was performed using miniscopes while following the guidelines on the miniscope.org website. Recording sessions were conducted on a plus-maze (with each arm being 70cm long and 7.5 cm wide) in which animals were trained to alternate between arms. A webcam mounted above the maze tracked the position of a green and red light emitting diode that were attached to the miniscope. These were used to determine position and head direction of the mouse. Images were acquired at 30Hz. All experimental procedures followed the guidelines approved by the McGill University Animal Care Committee.

The second dataset is from area CA1 of mouse hippocampus.
This dataset was collected using 2-photon imaging, while the head fixed male mouse was running for a stably placed hidden (non-cued) water reward on a 2 meter belt containing discrete tactile landmarks as in \cite{Zaremba2017}. Images were acquired at 60Hz (post hoc temporally decimated to 30Hz).

\subsection{Fitting encoding models to the data} 
\label{sec:model_fitting}
We use a similar maximum likelihood-based fitting procedure for both the simulated data and the two real datasets.  We denote $\theta=(\theta_1,\cdots,\theta_T)^{\top}\in (-180,180]$. We split the data into 60\% training data, 20\% validation data, and 20\% test data, for two simulation datasets and ADN data. For the CA1 data, only the data from running state are used, and the data are split it into 70\% training data, 10\% validation data, and 20\% test data.

\subparagraph{ZIG model}There are three parameters, i.e., the scale parameter $a$ and shape parameter $k$ for the gamma component, and the probability of non-zero responses $q$. We parameterize the scale $a$ and probability of non-zero responses $q$ as a function of stimulus $(\sin(\theta),\cos(\theta))$ using neural networks, i.e., \[\left(a_{t1},\cdots,a_{tN},q_{t1},\cdots,q_{tN}\right)=\left(f_1(\theta_t),\cdots,f_N(\theta_t),g_1(\theta_t),\cdots,g_N(\theta_t)\right),\]
where $f=(f_1,\cdots,f_N), g=(g_1,\cdots,g_N)$ are the output layer for $a$ and $q$ respectively. We use 2 hidden layers, each with tanh non-linearity, in the neural network.  For the output layer, we use a logistic link function for $f$ and an exponential link function for $g$. 30 nodes in hidden layers are used for the two simulated datasets and the ADN data; 15 nodes are used for the hippocampal data.  We fix the shape parameter $k$ to be a constant for individual neurons (i.e., $k$ is neuron-dependent but not $\theta$-dependent). 

We optimize all the parameters by maximizing the log-likelihood using a variant of gradient descent, i.e., Adam~\cite{kingma2014adam}. Specifically, the objective function can be expressed as
\begin{equation}
\begin{aligned}
    \arg\max\limits_{k,a,q}&\sum\limits_{i=1}^{N}\sum\limits_{t=1}^{T}\log\left(1-q_{ti}\right)\mathbb{1}(\hat{s}_{ti}=0)+\\
    &\left(\log q_{ti}+(k_i-1)\log(\hat{s}_{ti}-s_{min})-\frac{\hat{s}_{ti}-s_{min}}{a_{ti}}-k_i\log a_{ti}-\log\gamma(k_i)\right) \mathbb{1}(\hat{s}_{ti}>s_{min}).
\end{aligned}
\end{equation}

\subparagraph{Poisson model} We parameterize the Poisson mean $\lambda$ using a neural network with the same structure as described for the ZIG model above, with an exponential link function in the output layer.  Note that for the Poisson model, the likelihood function is not a proper likelihood because the the Poisson density can not be evaluated for non-integer values. We use a ``psuedo-likelihood'' function instead for the maximum likelihood estimation:
\begin{equation}
    \begin{aligned}
    \arg\max\limits_{\lambda}\sum\limits_{i=1}^{N}\sum\limits_{t=1}^{T}\hat{s}_{ti}\log\lambda_{ti}-\lambda_{ti}.
    \end{aligned}
\end{equation}

\subparagraph{Bernoulli model} The probability of positive response $q$ is parameterized using a neural network with the same structure as in the ZIG model, with a logistic link function in the output layer. Formally, the objective function can be defined as 
\begin{equation}
    \begin{aligned}
    \arg\max\limits_{q}\sum\limits_{i=1}^{N}\sum\limits_{t=1}^{T}(1-\hat{s}_{ti})\log\left(1-q_{ti}\right)+\hat{s}_{ti}\log q_{ti}.
    \end{aligned}
\end{equation}

\subparagraph{Gamma model} The scale parameter $a$ and the shape parameter $k$ are parameterized with the same neural network structure as in the ZIG model, except using an exponential link in the output layer. The objective function can be expressed as
\begin{equation}
\begin{aligned}
    \arg\max\limits_{k,a}&\sum\limits_{i=1}^{N}\sum\limits_{t=1}^{T}(k_i-1)\log\hat{s}_{ti}-\frac{\hat{s}_{ti}}{a_{ti}}-k_i\log a_{ti}-\log\gamma(k_i).
\end{aligned}
\end{equation}

\subsection{Shuffling analysis}
In section \ref{sec:real} we performed a shuffling analysis to investigate the conditional independence assumption used by the Bayesian decoder.  Details of this analysis are provided here.
We started with the original $T$-by-$N$ matrix $\hat{s}_{ti}$ (with $T$ denoting the number of observed video frames, and $N$ the number of extracted cells), then made a new matrix $\tilde{s}_{ti}$ as follows. For each time point $t$ and each cell $i$, we randomly chose a timestep $u$ ($u$ depends on $(t,i)$) such that $\theta_t=\theta_u$, and then set $\tilde{s}_{ti}=\hat{s}_{ui}$.
The new matrix $\tilde{s}$ has the same marginal distribution of $p(\hat{s}_t | \theta_t)$, so the encoding models will be the same, but the correlations between cells will be destroyed.

\section*{Acknowledgments}

We thank Tian Zheng and John Cunningham for helpful discussions. 

\section*{Conflict of interest}
The authors declare no conflicts of interest.


\bibliography{main}

\begin{thebibliography}{36}
\providecommand{\natexlab}[1]{#1}
\providecommand{\url}[1]{\texttt{#1}}
\providecommand{\urlprefix}{}

\bibitem[{Yuste and Denk(1995)Yuste, Rafael and Denk, Winfried}]{Yuste1995}
Yuste R, Denk W.
\newblock Dendritic spines as basic functional units of neuronal integration.
\newblock Nature 1995;375(6533):682.

\bibitem[{Svoboda et~al.(1997)Svoboda, Karel and Denk, Winfried and Kleinfeld,
  David and Tank, David W}]{Svoboda1997}
Svoboda K, Denk W, Kleinfeld D, Tank DW.
\newblock In vivo dendritic calcium dynamics in neocortical pyramidal neurons.
\newblock Nature 1997;385(6612):161.

\bibitem[{Helmchen et~al.(1999)Helmchen, Fritjof and Svoboda, Karel and Denk,
  Winfried and Tank, David W}]{Helmchen1999vivo}
Helmchen F, Svoboda K, Denk W, Tank DW.
\newblock In vivo dendritic calcium dynamics in deep-layer cortical pyramidal
  neurons.
\newblock Nature neuroscience 1999;2(11):989.

\bibitem[{Dombeck et~al.(2010)Dombeck, Daniel A and Harvey, Christopher D and
  Tian, Lin and Looger, Loren L and Tank, David W}]{Dombeck2010}
Dombeck DA, Harvey CD, Tian L, Looger LL, Tank DW.
\newblock Functional imaging of hippocampal place cells at cellular resolution
  during virtual navigation.
\newblock Nature neuroscience 2010;13(11):1433.

\bibitem[{Chen et~al.(2013)Chen, Tsai-Wen and Wardill, Trevor J and Sun, Yi and
  Pulver, Stefan R and Renninger, Sabine L and Baohan, Amy and Schreiter, Eric
  R and Kerr, Rex A and Orger, Michael B and Jayaraman, Vivek and
  others}]{Chen2013}
Chen TW, Wardill TJ, Sun Y, Pulver SR, Renninger SL, Baohan A, et~al.
\newblock Ultrasensitive fluorescent proteins for imaging neuronal activity.
\newblock Nature 2013;499(7458):295.

\bibitem[{Ziv et~al.(2013)Ziv, Yaniv and Burns, Laurie D and Cocker, Eric D and
  Hamel, Elizabeth O and Ghosh, Kunal K and Kitch, Lacey J and El Gamal, Abbas
  and Schnitzer, Mark J}]{Ziv2013}
Ziv Y, Burns LD, Cocker ED, Hamel EO, Ghosh KK, Kitch LJ, et~al.
\newblock Long-term dynamics of CA1 hippocampal place codes.
\newblock Nature neuroscience 2013;16(3):264.

\bibitem[{Rubin et~al.(2015)Rubin, Alon and Geva, Nitzan and Sheintuch, Liron
  and Ziv, Yaniv}]{Rubin2015}
Rubin A, Geva N, Sheintuch L, Ziv Y.
\newblock Hippocampal ensemble dynamics timestamp events in long-term memory.
\newblock Elife 2015;4:e12247.

\bibitem[{Driscoll et~al.(2017)Driscoll, Laura N and Pettit, Noah L and
  Minderer, Matthias and Chettih, Selmaan N and Harvey, Christopher
  D}]{Driscoll2017}
Driscoll LN, Pettit NL, Minderer M, Chettih SN, Harvey CD.
\newblock Dynamic reorganization of neuronal activity patterns in parietal
  cortex.
\newblock Cell 2017;170(5):986--999.

\bibitem[{Vogelstein et~al.(2009)Vogelstein, Joshua T and Watson, Brendon O and
  Packer, Adam M and Yuste, Rafael and Jedynak, Bruno and Paninski,
  Liam}]{Vogelstein2009}
Vogelstein JT, Watson BO, Packer AM, Yuste R, Jedynak B, Paninski L.
\newblock Spike inference from calcium imaging using sequential Monte Carlo
  methods.
\newblock Biophysical journal 2009;97(2):636--655.

\bibitem[{Vogelstein et~al.(2010)Vogelstein, Joshua T and Packer, Adam M and
  Machado, Timothy A and Sippy, Tanya and Babadi, Baktash and Yuste, Rafael and
  Paninski, Liam}]{Vogelstein2010}
Vogelstein JT, Packer AM, Machado TA, Sippy T, Babadi B, Yuste R, et~al.
\newblock Fast nonnegative deconvolution for spike train inference from
  population calcium imaging.
\newblock Journal of neurophysiology 2010;104(6):3691--3704.

\bibitem[{Pnevmatikakis et~al.(2016)Pnevmatikakis, Eftychios A and Soudry,
  Daniel and Gao, Yuanjun and Machado, Timothy A and Merel, Josh and Pfau,
  David and Reardon, Thomas and Mu, Yu and Lacefield, Clay and Yang, Weijian
  and others}]{Pnevmatikakis2016}
Pnevmatikakis EA, Soudry D, Gao Y, Machado TA, Merel J, Pfau D, et~al.
\newblock Simultaneous denoising, deconvolution, and demixing of calcium
  imaging data.
\newblock Neuron 2016;89(2):285--299.

\bibitem[{Deneux et~al.(2016)Deneux, Thomas and Kaszas, Attila and Szalay,
  Gergely and Katona, Gergely and Lakner, Tam{\'a}s and Grinvald, Amiram and
  R{\'o}zsa, Bal{\'a}zs and Vanzetta, Ivo}]{Deneux2016}
Deneux T, Kaszas A, Szalay G, Katona G, Lakner T, Grinvald A, et~al.
\newblock Accurate spike estimation from noisy calcium signals for ultrafast
  three-dimensional imaging of large neuronal populations in vivo.
\newblock Nature communications 2016;7:12190.

\bibitem[{Theis et~al.(2016)Theis, Lucas and Berens, Philipp and Froudarakis,
  Emmanouil and Reimer, Jacob and Ros{\'o}n, Miroslav Rom{\'a}n and Baden, Tom
  and Euler, Thomas and Tolias, Andreas S and Bethge, Matthias}]{Theis2016}
Theis L, Berens P, Froudarakis E, Reimer J, Ros{\'o}n MR, Baden T, et~al.
\newblock Benchmarking spike rate inference in population calcium imaging.
\newblock Neuron 2016;90(3):471--482.

\bibitem[{Friedrich et~al.(2017)Friedrich, Johannes and Zhou, Pengcheng and
  Paninski, Liam}]{Friedrich2017}
Friedrich J, Zhou P, Paninski L.
\newblock Fast online deconvolution of calcium imaging data.
\newblock PLoS computational biology 2017;13(3):e1005423.

\bibitem[{Jewell et~al.(2018)Jewell, Sean and Hocking, Toby Dylan and
  Fearnhead, Paul and Witten, Daniela}]{Jewell2018}
Jewell S, Hocking TD, Fearnhead P, Witten D.
\newblock Fast nonconvex deconvolution of calcium imaging data.
\newblock arXiv preprint arXiv:180207380 2018;.

\bibitem[{Speiser et~al.(2017)Speiser, Artur and Yan, Jinyao and Archer, Evan W
  and Buesing, Lars and Turaga, Srinivas C and Macke, Jakob H}]{Speiser2017}
Speiser A, Yan J, Archer EW, Buesing L, Turaga SC, Macke JH.
\newblock Fast amortized inference of neural activity from calcium imaging data
  with variational autoencoders.
\newblock In: Advances in Neural Information Processing Systems; 2017. p.
  4024--4034.

\bibitem[{Aitchison et~al.(2017)Aitchison, Laurence and Russell, Lloyd and
  Packer, Adam M and Yan, Jinyao and Castonguay, Philippe and Hausser, Michael
  and Turaga, Srinivas C}]{Aitchison2017}
Aitchison L, Russell L, Packer AM, Yan J, Castonguay P, Hausser M, et~al.
\newblock Model-based Bayesian inference of neural activity and connectivity
  from all-optical interrogation of a neural circuit.
\newblock In: Advances in Neural Information Processing Systems; 2017. p.
  3486--3495.

\bibitem[{Berens et~al.(2018)Berens, Philipp and Freeman, Jeremy and Deneux,
  Thomas and Chenkov, Nicolay and McColgan, Thomas and Speiser, Artur and
  Macke, Jakob H and Turaga, Srinivas C and Mineault, Patrick and Rupprecht,
  Peter and others}]{Berens2018}
Berens P, Freeman J, Deneux T, Chenkov N, McColgan T, Speiser A, et~al.
\newblock Community-based benchmarking improves spike rate inference from
  two-photon calcium imaging data.
\newblock PLoS computational biology 2018;14(5):e1006157.

\bibitem[{Pachitariu et~al.(2018)Pachitariu, Marius and Stringer, Carsen and
  Harris, Kenneth D}]{Pachitariu2018}
Pachitariu M, Stringer C, Harris KD.
\newblock Robustness of spike deconvolution for neuronal calcium imaging.
\newblock Journal of Neuroscience 2018;38(37):7976--7985.

\bibitem[{Greenberg et~al.(2018)Greenberg, David S and Wallace, Damian J and
  Voit, Kay-Michael and Wuertenberger, Silvia and Czubayko, Uwe and Monsees,
  Arne and Handa, Takashi and Vogelstein, Joshua T and Seifert, Reinhard and
  Groemping, Yvonne and others}]{Greenberg2018}
Greenberg DS, Wallace DJ, Voit KM, Wuertenberger S, Czubayko U, Monsees A,
  et~al.
\newblock Accurate action potential inference from a calcium sensor protein
  through biophysical modeling.
\newblock bioRxiv 2018;p. 479055.

\bibitem[{Stringer and Pachitariu(2019)Stringer, Carsen and Pachitariu,
  Marius}]{Stringer2019}
Stringer C, Pachitariu M.
\newblock Computational processing of neural recordings from calcium imaging
  data.
\newblock Current opinion in neurobiology 2019;55:22--31.

\bibitem[{Paninski and Cunningham(2018)Paninski, Liam and Cunningham, John
  P}]{paninski2018neural}
Paninski L, Cunningham JP.
\newblock Neural data science: accelerating the experiment-analysis-theory
  cycle in large-scale neuroscience.
\newblock Current opinion in neurobiology 2018;50:232--241.

\bibitem[{Jewell and Witten(2018)Jewell, Sean and Witten,
  Daniela}]{Jewell2018a}
Jewell S, Witten D.
\newblock Exact spike train inference via $\ell_0$ optimization.
\newblock The annals of applied statistics 2018;12(4):2457.

\bibitem[{L{\"u}tcke et~al.(2013)L{\"u}tcke, Henry and Gerhard, Felipe and
  Zenke, Friedemann and Gerstner, Wulfram and Helmchen, Fritjof}]{Lutcke2013}
L{\"u}tcke H, Gerhard F, Zenke F, Gerstner W, Helmchen F.
\newblock Inference of neuronal network spike dynamics and topology from
  calcium imaging data.
\newblock Frontiers in neural circuits 2013;7:201.

\bibitem[{Dana et~al.(2016)Dana, Hod and Mohar, Boaz and Sun, Yi and Narayan,
  Sujatha and Gordus, Andrew and Hasseman, Jeremy P and Tsegaye, Getahun and
  Holt, Graham T and Hu, Amy and Walpita, Deepika and others}]{Dana2016}
Dana H, Mohar B, Sun Y, Narayan S, Gordus A, Hasseman JP, et~al.
\newblock Sensitive red protein calcium indicators for imaging neural activity.
\newblock Elife 2016;5:e12727.

\bibitem[{Tomko and Crapper(1974)Tomko, George J and Crapper, Donald
  R}]{Tomko1974}
Tomko GJ, Crapper DR.
\newblock Neuronal variability: non-stationary responses to identical visual
  stimuli.
\newblock Brain research 1974;79(3):405--418.

\bibitem[{Tolhurst et~al.(1981)Tolhurst, DJ and Movshon, J Anthony and
  Thompson, ID}]{Tolhurst1981}
Tolhurst D, Movshon JA, Thompson I.
\newblock The dependence of response amplitude and variance of cat visual
  cortical neurones on stimulus contrast.
\newblock Experimental brain research 1981;41(3-4):414--419.

\bibitem[{Goris et~al.(2014)Goris, Robbe LT and Movshon, J Anthony and
  Simoncelli, Eero P}]{Goris2014}
Goris RL, Movshon JA, Simoncelli EP.
\newblock Partitioning neuronal variability.
\newblock Nature neuroscience 2014;17(6):858.

\bibitem[{Davidson et~al.(2009)Davidson, Thomas J and Kloosterman, Fabian and
  Wilson, Matthew A}]{Davidson2009}
Davidson TJ, Kloosterman F, Wilson MA.
\newblock Hippocampal replay of extended experience.
\newblock Neuron 2009;63(4):497--507.

\bibitem[{Ganmor et~al.(2016)Ganmor, Elad and Krumin, Michael and Rossi, Luigi
  F and Carandini, Matteo and Simoncelli, Eero P}]{Ganmor2016}
Ganmor E, Krumin M, Rossi LF, Carandini M, Simoncelli EP.
\newblock Direct estimation of firing rates from calcium imaging data.
\newblock arXiv preprint arXiv:160100364 2016;.

\bibitem[{Triplett et~al.(2019)Triplett, Marcus A and Pujic, Zac and Sun, Biao
  and Avitan, Lilach and Goodhill, Geoffrey J}]{triplett2019model}
Triplett MA, Pujic Z, Sun B, Avitan L, Goodhill GJ.
\newblock Model-based decoupling of evoked and spontaneous neural activity in
  calcium imaging data.
\newblock bioRxiv 2019;p. 691261.

\bibitem[{Picardo et~al.(2016)Picardo, Michel A and Merel, Josh and Katlowitz,
  Kalman A and Vallentin, Daniela and Okobi, Daniel E and Benezra, Sam E and
  Clary, Rachel C and Pnevmatikakis, Eftychios A and Paninski, Liam and Long,
  Michael A}]{picardo2016population}
Picardo MA, Merel J, Katlowitz KA, Vallentin D, Okobi DE, Benezra SE, et~al.
\newblock Population-level representation of a temporal sequence underlying
  song production in the zebra finch.
\newblock Neuron 2016;90(4):866--876.

\bibitem[{Mishchenko et~al.(2011)Mishchenko, Yuriy and Vogelstein, Joshua T and
  Paninski, Liam and others}]{mishchenko2011bayesian}
Mishchenko Y, Vogelstein JT, Paninski L, et~al.
\newblock A Bayesian approach for inferring neuronal connectivity from calcium
  fluorescent imaging data.
\newblock The Annals of Applied Statistics 2011;5(2B):1229--1261.

\bibitem[{Zhou et~al.(2018)Zhou, Pengcheng and Resendez, Shanna L and
  Rodriguez-Romaguera, Jose and Jimenez, Jessica C and Neufeld, Shay Q and
  Giovannucci, Andrea and Friedrich, Johannes and Pnevmatikakis, Eftychios A
  and Stuber, Garret D and Hen, Rene and others}]{zhou2018efficient}
Zhou P, Resendez SL, Rodriguez-Romaguera J, Jimenez JC, Neufeld SQ, Giovannucci
  A, et~al.
\newblock Efficient and accurate extraction of in vivo calcium signals from
  microendoscopic video data.
\newblock Elife 2018;7:e28728.

\bibitem[{Zaremba et~al.(2017)Zaremba, Jeffrey D and Diamantopoulou, Anastasia
  and Danielson, Nathan B and Grosmark, Andres D and Kaifosh, Patrick W and
  Bowler, John C and Liao, Zhenrui and Sparks, Fraser T and Gogos, Joseph A and
  Losonczy, Attila}]{Zaremba2017}
Zaremba JD, Diamantopoulou A, Danielson NB, Grosmark AD, Kaifosh PW, Bowler JC,
  et~al.
\newblock Impaired hippocampal place cell dynamics in a mouse model of the
  22q11. 2 deletion.
\newblock Nature neuroscience 2017;20(11):1612.

\bibitem[{Kingma and Ba(2014)Kingma, Diederik P and Ba, Jimmy}]{kingma2014adam}
Kingma DP, Ba J.
\newblock Adam: A method for stochastic optimization.
\newblock arXiv preprint arXiv:14126980 2014;.

\end{thebibliography}

\newpage
\section*{Supplementary information}
\beginsupplement

\end{document}